\documentstyle[11pt,aaspp4,flushrt]{article}

\slugcomment{\it Submitted to the Astrophysical Journal}
\lefthead{Stern et al.}
\righthead{Brown Dwarfs and High-Redshift Quasars}

\def\ie{{i.e.,~}}
\def\eg{{e.g.,~}}
\def\etal{{et al.,~}}

\def\deg{\ifmmode {^{\circ}}\else {$^\circ$}\fi}
\def\kms{\ifmmode {\rm\,km\,s^{-1}}\else
    ${\rm\,km\,s^{-1}}$\fi}
\def\ergcm2s{\ifmmode {\rm\,ergs\,cm^{-2}\,s^{-1}}\else
    ${\rm\,ergs\,cm^{-2}\,s^{-1}}$\fi}
\def\ergAcm2s{\ifmmode {\rm\,ergs\,cm^{-2}\,s^{-1}\,\AA^{-1}}\else
    ${\rm\,ergs\,cm^{-2}\,s^{-1}\,\AA^{-1}}$\fi}
\def\ergs{\ifmmode {\rm\,ergs\,s^{-1}}\else
    ${\rm\,ergs\,s^{-1}}$\fi}
\def\kmsMpc{\ifmmode {\rm\,km\,s^{-1}\,Mpc^{-1}}\else
    ${\rm\,km\,s^{-1}\,Mpc^{-1}}$\fi}

\def\spose#1{\hbox to 0pt{#1\hss}}
\def\simlt{\mathrel{\spose{\lower 3pt\hbox{$\mathchar"218$}}
     \raise 2.0pt\hbox{$\mathchar"13C$}}}
\def\simgt{\mathrel{\spose{\lower 3pt\hbox{$\mathchar"218$}}
     \raise 2.0pt\hbox{$\mathchar"13E$}}}

\def\plotfiddle#1#2#3#4#5#6#7{\centering \leavevmode
\vbox to#2{\rule{0pt}{#2}}
\includegraphics{#1}}

\def\tdwarf{ISS~J142950.9+333012}
\def\qso{ISS~J142738.5+331242}

\begin{document}

\title{Mid-Infrared Selection of Brown Dwarfs and High-Redshift Quasars}

\author{Daniel Stern\altaffilmark{1}, 
J. Davy Kirkpatrick\altaffilmark{2}, 
Lori E. Allen\altaffilmark{3},
Chao Bian\altaffilmark{4},
Andrew Blain\altaffilmark{4},
Kate Brand\altaffilmark{5},
Mark Brodwin\altaffilmark{1}, 
Michael J. I. Brown\altaffilmark{6},
Richard Cool\altaffilmark{7},
Vandana Desai\altaffilmark{4},
Arjun Dey\altaffilmark{8}, 
Peter Eisenhardt\altaffilmark{1},
Anthony Gonzalez\altaffilmark{9},
Buell T. Jannuzi\altaffilmark{8}, 
Karin Menendez-Delmestre\altaffilmark{4}, 
Howard A. Smith\altaffilmark{3}, 
B. T. Soifer\altaffilmark{4,10},
Glenn P. Tiede\altaffilmark{11} \&
E. Wright\altaffilmark{12}}

\altaffiltext{1}{Jet Propulsion Laboratory, California Institute of
Technology, 4800 Oak Grove Dr., Mail Stop 169-506, Pasadena, CA
91109 [e-mail: {\tt stern@zwolfkinder.jpl.nasa.gov}]}

\altaffiltext{2}{Infrared Processing and Analysis Center, California
Institute of Technology, Pasadena, CA 91125}

\altaffiltext{3}{Harvard-Smithsonian Center for Astrophysics, 60 Garden
St., Cambridge, MA 02138}

\altaffiltext{4}{Division of Physics, Math, and Astronomy, California
Institute of Technology, Pasadena, CA 91125}

\altaffiltext{5}{Space Telescope Science Institute, 3700 San Martin Dr.,
Baltimore, MD 21218}

\altaffiltext{6}{Princeton University Observatory, Peyton Hall, Princeton
University, Princeton, NJ 08544}

\altaffiltext{7}{Steward Observatory, University of Arizona, 933 N.
Cherry Ave., Tucson, AZ 85721}

\altaffiltext{8}{National Optical Astronomy Observatory, 950 N. Cherry
Ave., Tucson, AZ 85719}

\altaffiltext{9}{Department of Astronomy, University of Florida,
Gainesville, FL 32611}

\altaffiltext{10}{{\it Spitzer} Science Center, California Institute of
Technology, Pasadena, CA 91125}

\altaffiltext{11}{Department of Physics and Astronomy, Bowling Green
State University, Bowling Green, OH 43403}

\altaffiltext{12}{Department of Physics and Astronomy, University of
California at Los Angeles, Los Angeles, CA 90095}

\begin{abstract} 

We discuss color selection of rare objects in a wide-field, multiband
survey spanning from the optical to the mid-infrared.  Simple color
criteria simultaneously identify and distinguish two of the most sought
after astrophysical sources:  the coolest brown dwarfs and the most
distant quasars.  We present spectroscopically-confirmed examples of
each class identified in the IRAC Shallow Survey of the Bo\"otes field
of the NOAO Deep Wide-Field Survey.  \tdwarf\ is a T4.5 brown dwarf at
a distance of approximately 42~pc, and \qso\ is a radio-loud quasar at
redshift $z = 6.12$.  Our selection criteria identify a total of four
candidates over 8 square degrees of the Bo\"otes field.  The other two
candidates are both confirmed $5.5 < z < 6$ quasars, previously reported
by Cool et al. (2006).  We discuss the implications of these discoveries
and conclude that there are excellent prospects for extending such
searches to cooler brown dwarfs and higher redshift quasars.

\end{abstract}


\section{Introduction}

Wide-area surveys are one of the most powerful tools for observational
astronomy, and have led to discoveries ranging from Earth-crossing
asteroids to the most distant quasars.  Historically, when technology
allows a wavelength regime to be newly probed, either in terms of
sensitivity or area, one of the first tasks is a large, shallow survey
to see what astrophysical phenomena lurk in the uncovered territory.
In recent years, major advances from this line of research include the
discovery of the coolest Galactic stars by the Two Micron All Sky Survey
(2MASS), ultraluminous infrared galaxies by the {\it Infrared Astronomical
Satellite}, the most distant quasars by the Sloan Digital Sky Survey
(SDSS), and the power spectrum of the cosmic microwave background,
first by the {\it Cosmic Background Explorer} and later refined by the
{\it Wilkinson Microwave Anisotropy Probe}.  Such fundamental scientific
discoveries have been a major incentive and reward for NASA's Explorer
program and other large projects.  The pace of scientific discovery
relies on such programs continuing.

The mid-infrared regime has been made newly accessible by the launch
of the {\it Spitzer Space Telescope} \markcite{Werner:04}(Werner {et~al.} 2004).  At its least
competitive, shortest waveband, $3.6 \mu$m, {\it Spitzer} is still
more than five orders of magnitude more efficient than the largest
ground-based observatories for areal surveys.  For the longest wavebands,
ground-based observations are simply not possible.  Even compared to
previous space-based missions, {\it Spitzer} offers several orders of
magnitude increase in mapping efficiency.

Thus inspired, we have undertaken a shallow, wide-area 3.6 to $8.0~\mu$m
survey with {\it Spitzer}, summarized in \S2.  We discuss two of the
rare, interesting astronomical sources which are ideally suited to
selection by combining deep optical data with shallow mid-infrared
data:  the coolest Galactic brown dwarfs and the most distant quasars.
The former, of course, are not actually rare in the cosmos; their
faint optical magnitudes merely delayed their discovery until
recent years and continue to make them ``rare'' in terms of known,
spectroscopically-confirmed examples.  Section~3 discusses the selection
criteria used to identify such sources and \S4 describes our spectroscopic
observations which confirmed both a cool brown dwarf (spectral class
T4.5) and a high-redshift ($z = 6.12$) quasar.  The implications for
these discoveries are described in \S4, and \S5 summarizes the results
and discusses future prospects.  Throughout we adopt a ($\Omega_m,
\Omega_\Lambda$) = ($0.3, 0.7$) flat cosmology and $H_0 = 70 \kmsMpc$.
Unless otherwise stated, all magnitudes are quoted in the Vega system.

\section{Multiwavelength Surveys of Bo\"otes}

The Bo\"otes field is a 9 deg$^2$ field which has been the target of deep
observations across the electromagnetic spectrum.  Bo\"otes was initially
selected as the North Galactic field of the NOAO Deep Wide-Field Survey
\markcite{Jannuzi:99}(NDWFS; Jannuzi \& Dey 1999), which obtained deep optical ($B_WRI$)
and moderately-deep near-infrared ($K_s$) images across the entire
field.\footnote{The third data release is publicly available at {\tt
http://www.archive.noao.edu/ndwfs}.  The optical filter specifications
are $B_W$ ($\lambda_c = 4111$~\AA, FWHM $= 1275$~\AA), $R$ ($\lambda_c
= 6514$~\AA, FWHM $= 1511$~\AA), and $I$ ($\lambda_c = 8205$~\AA, FWHM
$= 1915$~\AA).  The $R$ and $I$ filters are part of the Harris filter
set and the photometry was calibrated to match the Cousins system.
The near-infrared filter conforms to the standard filter set.}  These
images reach approximate 5$\sigma$ point source depths of $B_W = 27.1$,
$R = 26.1$, $I = 25.4$ (B.~Jannuzi \etal in prep.)  and $K_s = 19.0$
(A.~Dey \etal in prep.).  Subsequently, the field has been observed at
X-ray energies with the {\it Chandra X-Ray Observatory} \markcite{Murray:05}(Murray {et~al.} 2005),
with a $z'$ filter using the Bok 2.3m telescope at Kitt Peak (R.~Cool
in prep.), more deeply in the near-infrared ($JK_s$) as part of the
FLAMINGOS Extragalactic Survey \markcite{Elston:06}(FLAMEX; Elston {et~al.} 2006), in the
infrared with the {\it Spitzer Space Telescope} \markcite{Eisenhardt:04,
Papovich:04}(Eisenhardt {et~al.} 2004; Papovich {et~al.} 2004), and at radio frequencies using the Westerbork Synthesis
Radio Telescope \markcite{deVries:02}(1.4~GHz; {de~Vries} {et~al.} 2002) and the Very Large Array
(325~MHz; S.~Croft \etal in prep.).  Approximately 20,000 spectroscopic
redshifts in the Bo\"otes field have been obtained by the AGN and Galaxy
Evolution Survey (AGES; C.~Kochanek \etal in prep.) and \markcite{Brodwin:06}Brodwin {et~al.} (2006)
reports on nearly 200,000 photometric redshifts in this field.

The mid-infrared imaging of Bo\"otes is central to this paper.  As part
of a guaranteed-time observation program, 8 deg$^2$ of the field was
imaged with the Infrared Array Camera \markcite{Fazio:04a}(IRAC; Fazio {et~al.} 2004) at
3.6 to 8 $\mu$m.  \markcite{Eisenhardt:04}Eisenhardt {et~al.} (2004) presents the survey design,
reduction, calibration, and initial results.  The survey, called the
IRAC Shallow Survey, identifies $\approx$ 270,000, 200,000, 27,000, and
26,000 sources brighter than $5\sigma$ Vega magnitude limits of 18.4,
17.7, 15.5, and 14.8 at 3.6, 4.5, 5.8, and 8.0 $\mu$m, respectively,
where IRAC magnitudes are measured in 6\arcsec\ diameter apertures and
corrected to total magnitudes assuming sources are unresolved at the
1\farcs66 $-$ 1\farcs98 resolution of IRAC.

\begin{figure}[!t] 
\begin{center} 
\plotfiddle{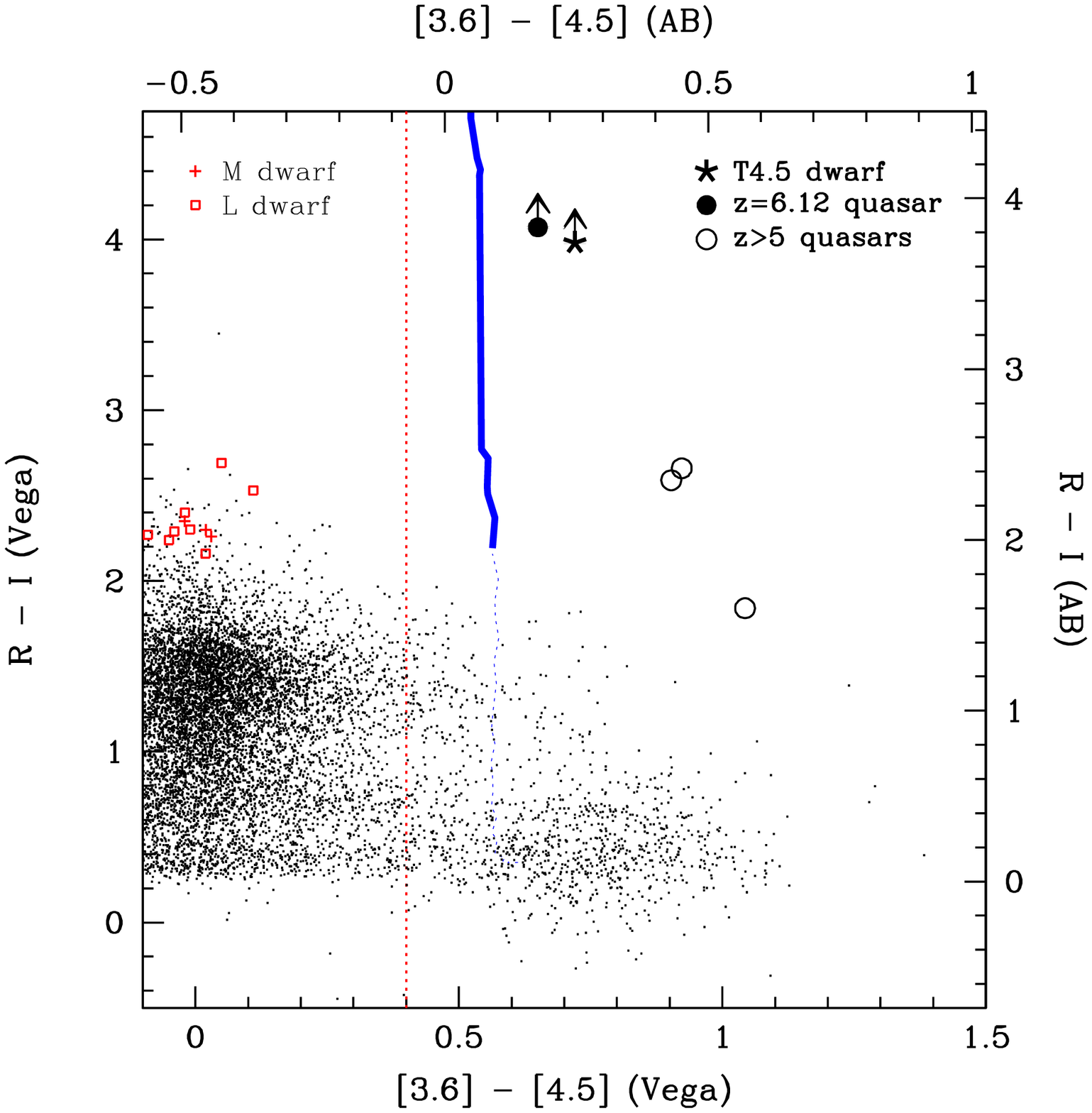}{2.0in}{0}{40}{40}{-240}{-100}
\plotfiddle{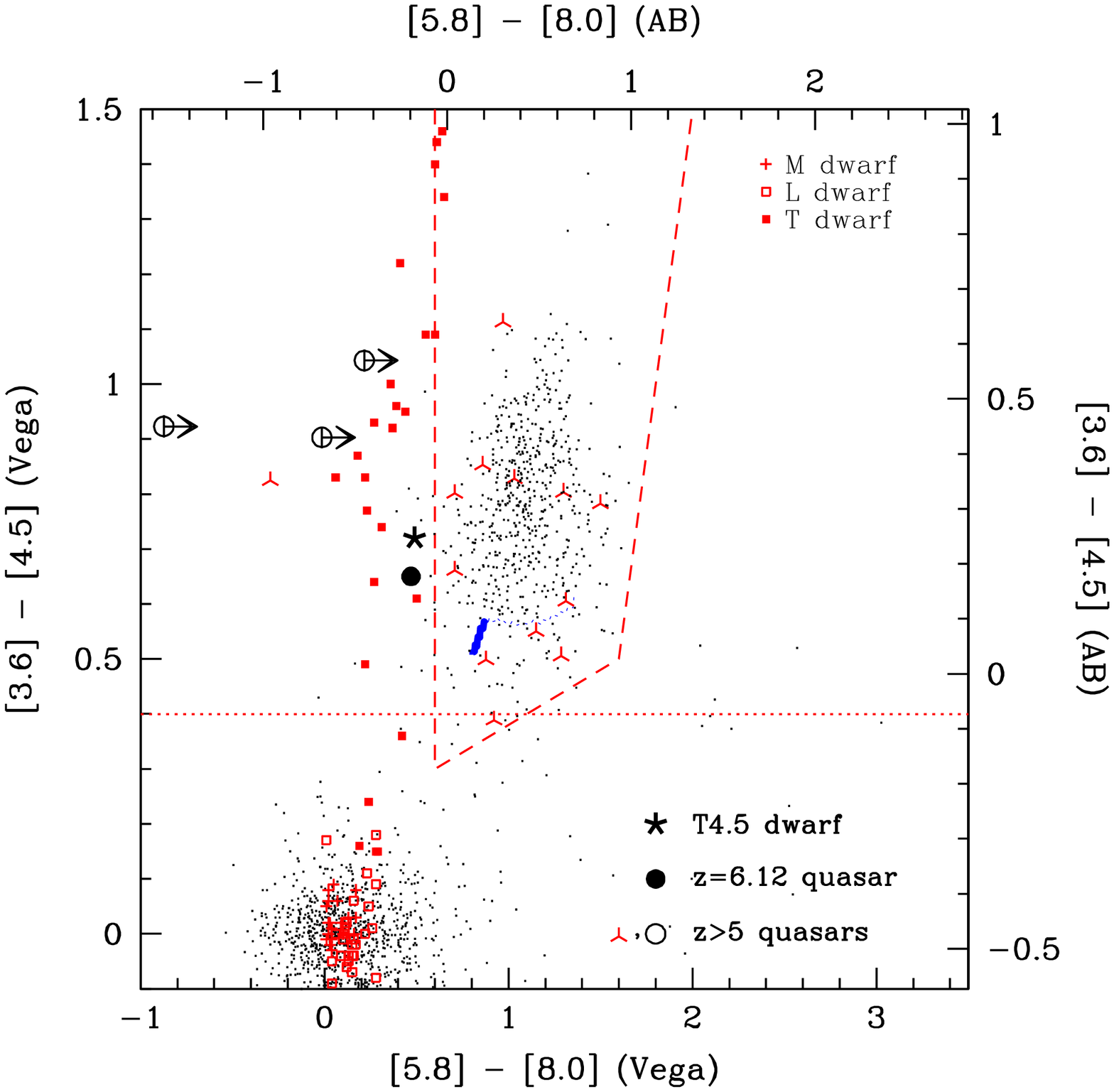}{0.0in}{0}{40}{40}{0}{-70}
\end{center}

\caption{Color-color diagrams for unresolved sources in the Bo\"otes
field.  As indicated, the asterisk refers to \tdwarf\ (T4.5 brown
dwarf), the filled circle refers to \qso\ ($z = 6.12$ quasar), and the
open circles refer to Bo\"otes field $5 < z < 6$ quasars from Cool
et al. (2006).  The dashed line in the right panel illustrates the
empirically determined wedge largely populated by luminous, unobscured
AGN (Stern et al. 2005).  The dotted line illustrates the selection
criteria employed in Cool et al. (2006) to identify luminous AGN which are
too faint to be detected in all IRAC channels, [3.6] $-$ [4.5] $>$ 0.4.
Dots illustrate typical colors of sources which are unresolved in the $I$
band (stellarity index $\geq 0.8$).  In the left panel, sources with $18 <
I < 20$ and $\geq 5 \sigma$ detections in [3.6] and [4.5] are plotted.
In the right panel, sources with $10 < I < 20$ and $\geq 5 \sigma$
detections in all four IRAC passbands are plotted.  Photometry of M, L,
and T dwarfs from Dahn et al. (2002; optical) and Patten et al. (2006;
IRAC) are plotted in red, as indicated.  SDSS quasars at $z \approx 6$
are plotted as inverted-Y's (Jiang et al. 2006).  Quasars and typical
stars are clearly separated on the basis of their mid-infrared colors.
The blue line illustrates the colors of the SDSS quasar template
from Richards et al. (2006) for $3 \leq z \leq 7$, subject to the Madau
(1995) formulation for the opacity of the intergalactic medium as a
function of redshift.  The line becomes thicker for $z \geq 5.5$.}


\label{fig.colcol}
\end{figure}

\begin{figure}[!t]
\begin{center}
\plotfiddle{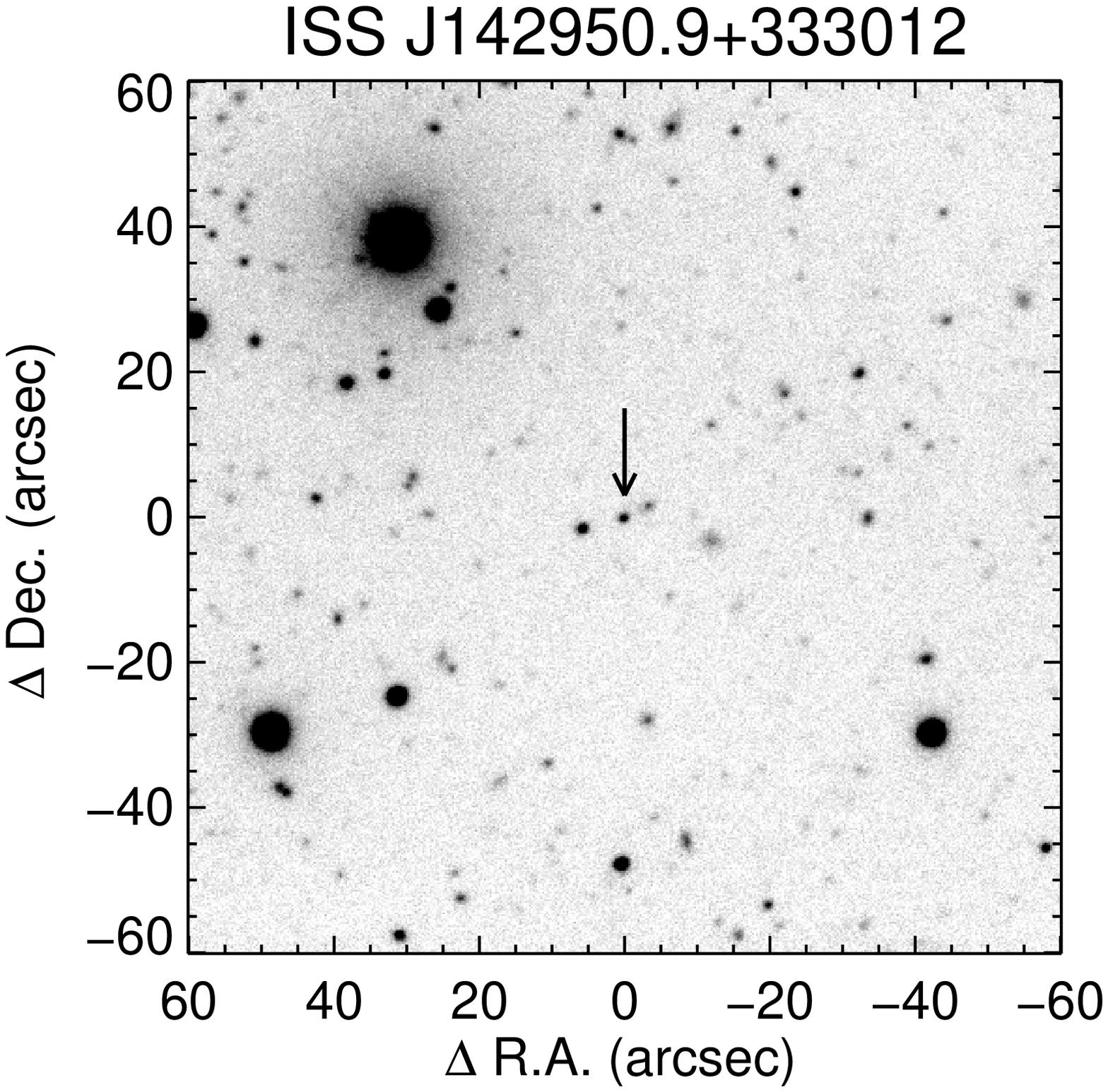}{1.9in}{0}{45}{45}{-270}{-50}
\plotfiddle{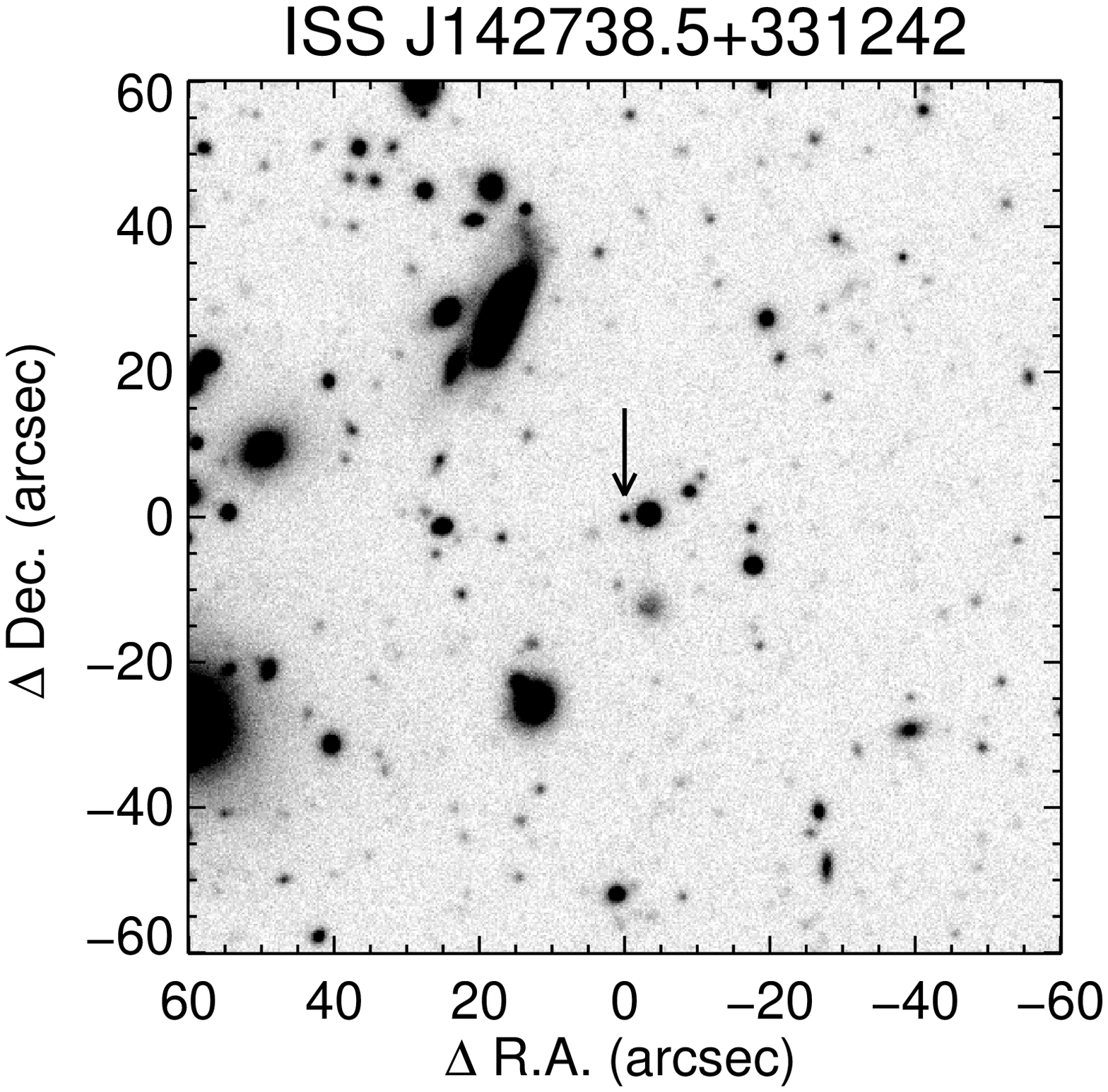}{0.0in}{0}{45}{45}{-20}{-20}
\end{center}

\caption{Finding charts for \tdwarf\ (T4.5 brown dwarf) and \qso\
($z=6.12$ quasar) from the NDWFS $I$-band imaging.  The fields are
2\arcmin\ $\times$ 2\arcmin, centered on the targets.  North is at the
top, and east is to the left.}

\label{fig.finders}
\end{figure}

\section{Mid-Infrared Selection of Rare Sources}

Two core science goals of the IRAC Shallow Survey are the identification
of the coolest stars and the identification of the most distant quasars.
Both are optically-faint sources that are difficult to find, but are
highly sought after for their astrophysical significance.  Brown dwarfs
probe the stellar-to-planetary link \markcite{Kirkpatrick:05}(\eg Kirkpatrick 2005),
while the highest redshift quasars probe the conditions of the early
universe and the onset of cosmic reionization \markcite{Fan:06}(\eg Fan {et~al.} 2006).
Currently, there are less than 100 T-type brown dwarfs known and only
ten quasars at $z > 6$.  Both brown dwarfs and the most distant quasars
are substantially brighter at mid-infrared wavelengths than at optical
wavelengths.  Therefore, wide-area, shallow, infrared surveys are ideally
suited to identifying samples of both types of sources.


Brown dwarfs have red colors due to their cool temperatures.
At mid-infrared wavelengths the spectra of {\it most} stars generally
follow a Rayleigh-Jeans tail, giving them mid-infrared Vega colors
near zero.  For cooler stars and brown dwarfs, however, the presence of
deep molecular absorptions results in very different emergent spectra
\markcite{Kirkpatrick:05}(Kirkpatrick 2005).  Specifically, the fundamental bands of CH$_4$
and CO between 3 and 5~$\mu$m \markcite{Oppenheimer:98, Cushing:05}(Oppenheimer {et~al.} 1998; Rayner \& Vacca 2005)
and additional bands of H$_2$O, CH$_4$, and NH$_3$ between 5 and 12
$\mu$m \markcite{Roellig:04}(Roellig {et~al.} 2004) dramatically recarve the spectral energy
distributions (SEDs) of these objects and give them unique IRAC colors
\markcite{Patten:06}(Patten {et~al.} 2006).  Shortward of 3~$\mu$m, H$_2$O bands in L and T
dwarfs (and CH$_4$ bands in T dwarfs only) cause deep depressions in the
near-infrared spectra \markcite{Mclean:03}(\eg McLean {et~al.} 2003), and pressure-broadened
\ion{Na}{1} and \ion{K}{1} resonance doublets suppress much of the
flux below 1~$\mu$m \markcite{Kirkpatrick:99, Burgasser:03}(Kirkpatrick {et~al.} 1999; Burgasser {et~al.} 2003a), making brown
dwarfs extremely faint in the optical.  Specifically, the colors of known
brown dwarfs later than type mid-T are $R-I>3.5$ \markcite{Kirkpatrick:99,
Dahn:02}(Kirkpatrick {et~al.} 1999; Dahn {et~al.} 2002), $0.7 <$ [3.6] $-$ [4.5] $< 2$, and $0 <$ [5.8] $-$ [8.0] $<
0.8$ \markcite{Patten:06}(Patten {et~al.} 2006).

High-redshift quasars have red colors primarily due to absorption by
foreground neutral hydrogen in the intergalactic medium which strongly
suppresses the intrinsic UV emission of these AGN.  At the highest
redshifts, $z \simgt 6$, very little flux is detectable below Ly$\alpha$,
providing the highest redshift quasars with similar optical colors to cool
stars.  Longward of Ly$\alpha$, luminous quasars are well-approximated
by a power law and are easily identified in mid-infrared color-color
diagrams \markcite{Stern:05b}(\eg Stern {et~al.} 2005).

Therefore, both the coolest brown dwarfs and the highest redshift
quasars should easily be identifiable by selecting unresolved sources
with very red optical colors and relatively flat (in $f_\nu$)
mid-infrared SEDs.  Fig.~\ref{fig.colcol} illustrates these selection
criteria for the Bo\"otes field.  Note that at the highest redshifts,
$z > 7$, quasars drop out of the optical completely.  Typical colors of
optical ($I$-band) point sources are presented.  These color-color
plots clearly separate stars and quasars, at least for typical, hot
stars and typical, moderate-redshift quasars:  most stars have
mid-infrared colors near zero, while quasars are distinguished by their
redder mid-infrared colors.  Three confirmed quasars in this field at
$5.39 \leq z \leq 5.85$ identified by \markcite{Cool:06}Cool {et~al.}
(2006) are indicated, as are twelve $z \approx 6$ quasars from the SDSS
\markcite{Jiang:06}(Jiang {et~al.} 2006).  The IRAC color-color
criteria empirically determined by \markcite{Stern:05b}Stern {et~al.}
(2005) and \markcite{Cool:06}Cool {et~al.} (2006) to select luminous
AGN are indicated.  Additionally, the colors of M, L, and T dwarfs from
\markcite{Dahn:02}Dahn {et~al.} (2002) and \markcite{Patten:06}Patten
{et~al.} (2006) are plotted.  The thick solid line shows the expected
colors of $3 \leq z \leq 7$ quasars, calculated using the
\markcite{Richards:06}Richards {et~al.} (2006) SDSS quasar template
subject to the \markcite{Madau:95}Madau (1995) formulation for the
opacity of the intergalactic medium as a function of redshift.  As can
be seen, the Ly$\alpha$ forest causes high-redshift quasars to become
very red in $R - I$ at $z \simgt 5$, while the mid-infrared colors vary
only slightly over this large redshift range.  Cool stars and
high-redshift quasars are identifiable from their red $R - I$ and [3.6]
$-$ [4.5] colors.  Longer-wavelength, [5.8] $-$ [8.0] colors can
provide additional information, but require deeper data to obtain
robust detections in these less-sensitive passbands.

As seen in Fig.~\ref{fig.colcol}, both brown dwarfs and high-redshift
quasars should be easily identified using the simple (Vega-system)
selection criteria of (i) $R - I \geq 2.5$, (ii) [3.6] $-$ [4.5] $\geq$
0.4, and (iii) unresolved at $I$-band.  To restrict the number of
spurious sources identified in the catalogs, we also require (iv) $B_W -
I \geq 2.5$.  These constraints implicitly require robust detections
(or robust non-detections) in the various bands.  In particular, the
$I$-band morphology criterion requires $I \simlt 23$ for the NDWFS survey.
According to the work of \markcite{Patten:06}Patten {et~al.} (2006), the IRAC color criterion
eliminates sources hotter than spectral class T3.  The Bo\"otes $4.5 \mu$m
catalog (ver.~1.3) identifies 30 candidates matching these selection
criteria, which are trimmed to four robust candidates after visual
inspection.  The most common cause of a false positive is source blending.
One source, ISS~J142918.1+343731, appeared modestly robust after visually
inspecting the ground-based imaging.  However, the source resides near a
$z > 1$ galaxy cluster identified by \markcite{Eisenhardt:06}Eisenhardt {et~al.} (2006).  {\it Hubble
Space Telescope} imaging of the cluster (GO~10836; P.I. S.~Perlmutter)
shows that the potential candidate is compact, but clearly resolved,
and thus unlikely to be either a brown dwarf or a high-redshift quasar.
Two of the final four candidates have already been spectroscopically
confirmed as $5.5 < z < 6$ quasars by \markcite{Cool:06}Cool {et~al.} (2006).  The remaining
two were targeted spectroscopically during Spring 2006, as discussed next.
Table~\ref{tab.bootes} presents all four Bo\"otes field candidates,
in order of decreasing $3.6 \mu$m flux, and Fig.~\ref{fig.finders}
presents finding charts for the two sources described in \S4.


\section{Spectroscopic Observations and Discussion}

Initial spectroscopic follow-up of candidates was obtained with the
Multi-Aperture Red Spectrometer \markcite{Barden:01}(MARS; Barden {et~al.} 2001) on the Mayall
4m telescope at Kitt Peak.  MARS is an optical spectrograph which uses
a high resistivity, p-channel Lawrence Berkeley National Laboratory
CCD with little fringing and very high throughput at long wavelengths
($\simlt 10,500$ \AA).  On the nights of UT 2006 March 24 $-$ 26,
we obtained spectra of red sources in the Bo\"otes field using the
1\farcs7 wide long slit, OG550 order-sorting filter, and the VG8050 grism.
Across much of the optical window, the instrument configuration provides
resolution $R \approx 1100$ spectra, as measured from sky lines filling
the slit.  \tdwarf\ was observed for 1.5~hr on UT 2006 March 24, split
into three dithered 1800~s exposures.  \qso\ was observed for 1~hr on
UT 2006 March 25, split into three 1200~s exposures.  The data were
processed following standard optical, slit spectroscopy procedures.
The nights were not photometric, but relative flux calibration of
the spectra was achieved with observations of the spectrophotometric
standards Feige~34 and PG~0823+546 \markcite{Massey:90}(Massey \& Gronwall 1990) obtained during the
same observing run.  The extracted, calibrated MARS spectra are presented
in Figs.~\ref{fig.tdwarfopt} and \ref{fig.qso}.  The bright star 3\farcs3
east of \qso\ made extraction of the fainter target challenging, resulting
in systematic fluctuations of the background at the 1~$\mu$Jy level.

Near-infrared spectroscopy of \tdwarf\ was obtained with the
cryogenic, cross-dispersed  Near-Infrared Echelle Spectrograph
\markcite{Mclean:98}(NIRSPEC; McLean {et~al.} 1998) on the Keck~II 10m telescope atop Mauna Kea.
We first obtained $J$- and $H$-band spectroscopy on UT 2006 April 05.
An AB nod sequence with a total on-source integration time of 200~s per
grating setting was used.  For both grating settings, the G2~V star GSPC
P300-E from \markcite{Colina:97}Colina \& Bohlin (1997) was used for both telluric correction and
flux calibration.  An additional $J$-band spectrum was acquired on UT 2006
May 11.  On this night a 300 s integration was taken both on-source
and off-source, and the F0 star BD+66 1089 was acquired for telluric
correction and flux calibration.  Fig.~\ref{fig.tdwarf} presents the
combined near-infrared spectrum.



\subsection{\tdwarf:  Mid-T Brown Dwarf}

\begin{figure}[!t]
\begin{center}
\plotfiddle{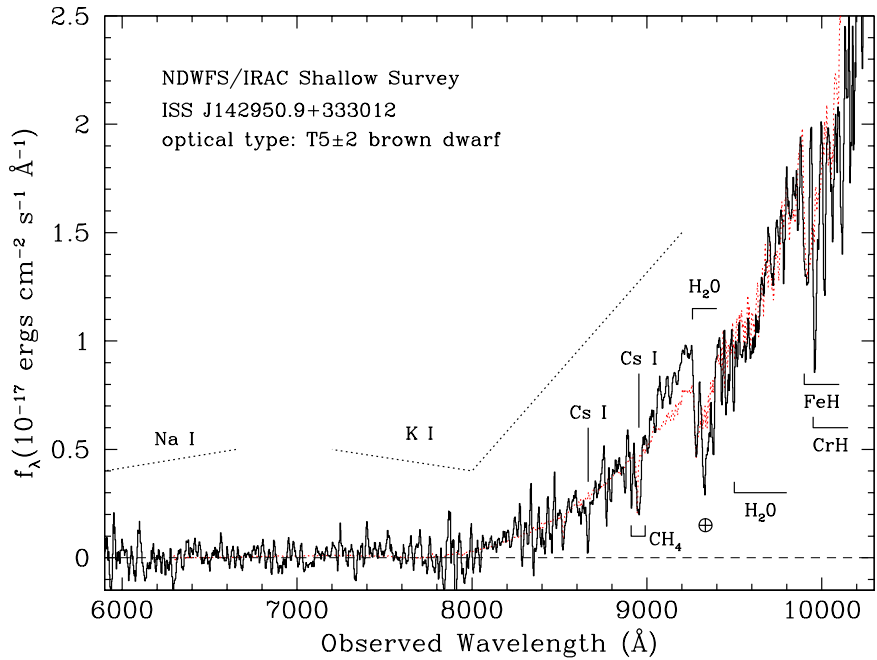}{3.1in}{-90}{45}{45}{-180}{260}
\end{center}

\caption{Optical spectrum of \tdwarf, optically classified as a T5$\pm$2
brown dwarf, obtained with the MARS spectrograph on KPNO~4m telescope.
The relative flux calibration was determined from observations of
standard stars from the same observing runs with the same instrumental
configurations.  The spectrophotometric scale was estimated from the
imaging.  The dotted spectrum shows 2MASS~J055919.14$-$140448.8, classified as
a T5 brown dwarf at optical wavelengths (Burgasser et al. 2003a).}

\label{fig.tdwarfopt}
\end{figure}

\begin{figure}[!t]
\begin{center}
\plotfiddle{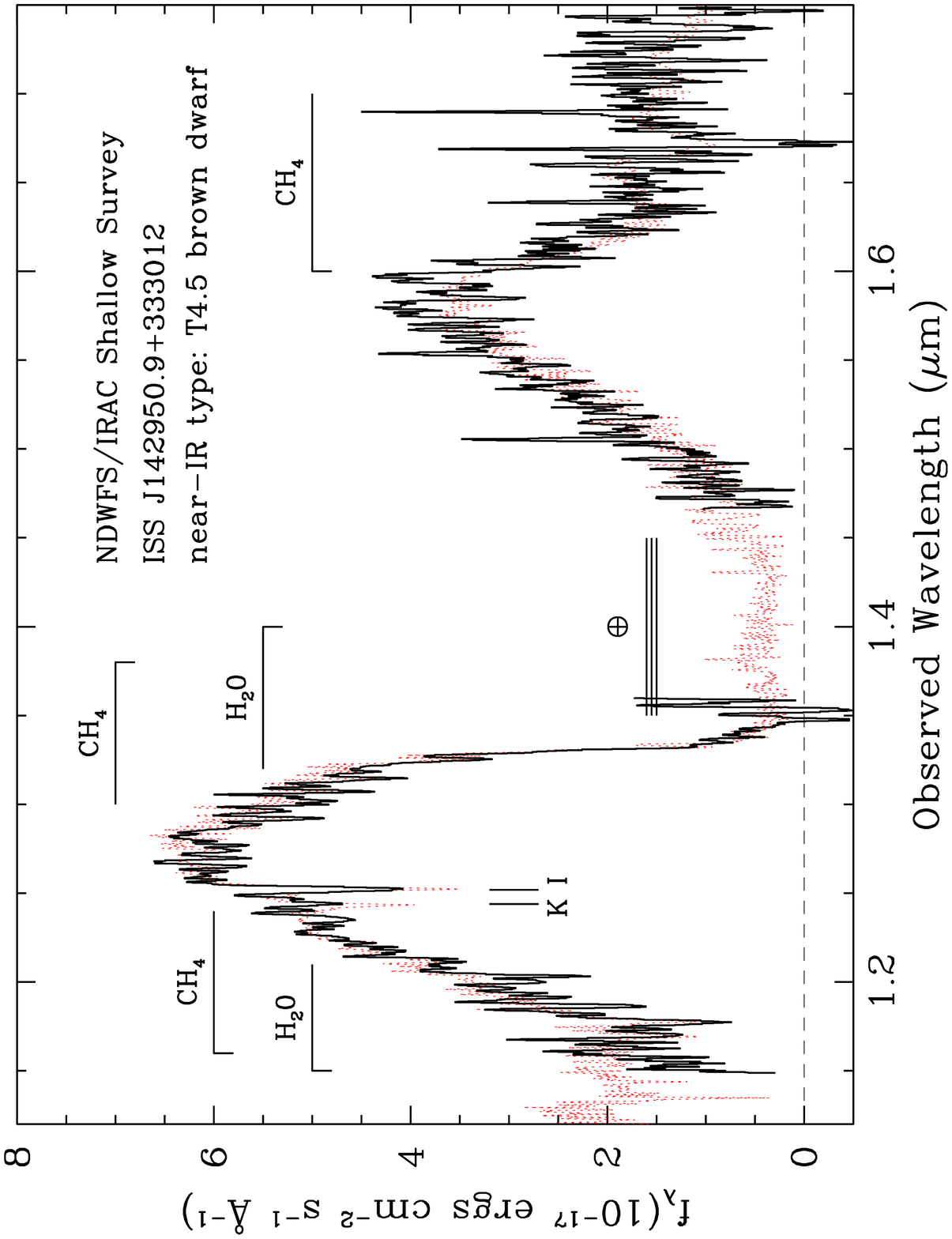}{3.1in}{-90}{45}{45}{-180}{260}
\end{center}

\caption{Near-infrared spectrum of \tdwarf\ obtained with the NIRSPEC
spectrograph on the Keck~II telescope, classified as a Galactic T4.5 brown
dwarf from these data.  The relative flux calibration was determined
from observations of standard stars from the same observing runs with
the same instrumental configurations.  The spectrophotometric scale
was estimated from the imaging.  The dotted line shows the infrared
spectrum of 2MASS~J0559$-$1404, classified as a T4.5 brown dwarf in the
near-infrared (McLean et al. 2003).}

\label{fig.tdwarf}
\end{figure}

The spectrum of \tdwarf\ shows the classic signatures of a T dwarf.
The optical spectrum in Fig.~\ref{fig.tdwarfopt} shows a sharp rise to
the longest wavelengths, indicative of a cool temperature and strong
absorption by the pressure-broadened wings of \ion{K}{1} (and to some
extent \ion{Na}{1}).  Even more telling are the $J$- and $H$-band spectra
in Fig.~\ref{fig.tdwarf} that show strong CH$_4$ and H$_2$O absorption,
the former of which is the hallmark of spectral class T.

As this object has both optical and near-infrared spectra, we can
classify on both the optical and near-infrared classification schemes.
The optical typing of T dwarfs is somewhat crude because the $\leq 1 \mu$m
spectra show less variation than at longer wavelengths.  Nonetheless,
\markcite{Burgasser:03}Burgasser {et~al.} (2003a) have established standards for classes T2, T5, T6,
and T8.  In the $6000 - 10000$ \AA\ range the best diagnostic is the
9300 \AA\ band of H$_2$O.  Unfortunately our MARS spectrum has not
been telluric corrected so the depth of this water feature will be
influenced by both the earth's atmosphere as well as the atmosphere
of the brown dwarf itself.  This feature in \tdwarf\ is not as deep
as in the spectrum of a T8, so the true spectral type must be earlier
than that.  Comparisons with the T2, T5, and T6 standards obtained with
Keck \markcite{Burgasser:03}(Burgasser {et~al.} 2003a) show that the overall slope most resembles that
of the T5.  Given the coarseness of classification in this wavelength
regime, we can assign only a crude optical spectral type of T5$\pm$2.

In the near-infrared the situation is much improved.  In this wavelength
regime there is a full set of standards for each spectral subtype
from T0 to T8 \markcite{Burgasser:06}(Burgasser {et~al.} 2006).  Using Keck NIRSPEC spectra from
\markcite{Mclean:03}McLean {et~al.} (2003) of the \markcite{Burgasser:06}Burgasser {et~al.} (2006) standards, we find that the
individual $J$-band spectra best match a type intermediate between T4 and
T5. A similar fit to the $H$-band data alone gives the identical result.
These results point to a solid near-infrared spectral type of T4.5.

Shown in Figs.~\ref{fig.tdwarfopt} - \ref{fig.tdwarf} are comparisons
of the spectra of \tdwarf\ and 2MASS~J0559$-$1404, which is the
optical T5 standard and typed as T4.5 on the \markcite{Burgasser:06}Burgasser {et~al.} (2006)
near-infrared scheme. (That is, 2MASS~J0559$-$1404 has the same type
as \tdwarf\ in both wavelength regimes.)  Note the similarities between
the two spectra.  2MASS~J0559$-$1404 has a well measured trigonometric
parallax of 97.7$\pm$1.3 mas \markcite{Dahn:02}(Dahn {et~al.} 2002) and an absolute magnitude
of $M_J = 13.75 \pm 0.04$, which allows us to estimate a distance
to \tdwarf\ of 42 pc, assuming both T dwarfs are single.  However,
2MASS~J0559$-$1404 is the most overluminous object in the early-/mid-T
``hump'' on the Hertzsprung-Russell diagram \markcite{Vrba:04,
Golimowski:04}(see Vrba {et~al.} 2004; Golimoski {et~al.} 2004), leading some researchers to believe that it might be
a close, equal-magnitude double despite all current evidence to the
contrary \markcite{Burgasser:03b, Gelino:05, Liu:06}(Burgasser {et~al.} 2003b; Gelino \& Kulkarni 2005; Liu {et~al.} 2006).  Correcting for this
possibility, we find that \tdwarf\ might be as close as 30 pc.  No other
optically classified T dwarfs are known of spectral type T5; the only
other T dwarf with a measured trigonometric parallax and near-infrared
type of T4.5 is SDSS~J020742.48+000056.2.  The parallax measurement of
34.85$\pm$9.87 mas for SDSS~J0207+0000 \markcite{Vrba:04}(Vrba {et~al.} 2004) implies $M_J =
14.51 \pm 0.64$, which is very uncertain but lends some weak support to
the closer distance estimate for \tdwarf.


The first images to detect the brown dwarf were the NDWFS $I$-band
observations obtained on UT 2000 April 28, 3.7~yr prior to the IRAC
imaging.  Comparing ten nearby sources detected in both the $I$-band
and $3.6 \mu$m observations, \tdwarf\ has a detected proper motion
of $0\farcs1 \pm 0\farcs03\, {\rm yr}^{-1}$, in a southerly direction.
This is comparable in amplitude to the expected reflex solar motion for a
source at $\approx 40$~pc.  Interestingly, the star 5\farcs7 east of the
brown dwarf shows a higher proper motion, $\mu = 0\farcs3 \pm 0\farcs03\,
{\rm yr}^{-1}$ in the NW direction.  The colors of this $R = 22.3$ star
(ISS~J142951.3+333010) are relatively blue, $B_W - R = 0.6, R-I = 0.6$,
suggesting a relatively hot white dwarf at a distance of several hundred
pc, moving at several hundred km sec$^{-1}$.

The $4.5 \mu$m flux of the brown dwarf is 2.7~mag brighter than the survey
limit (\ie $V/V_{\rm max} = 0.024$), whereas the $I$ magnitude is only
0.9~mag above the limit ($V/V_{\rm max} = 0.3$).  This suggests that the
$I < 23$ requirement imposed to provide robust morphological selection
of unresolved sources is a significant limiting factor.  We estimate
that our selection criteria restrict our sensitivity to brown dwarfs
of spectral type T3 to T6.  The former limit comes from the IRAC color
criterion \markcite{Patten:06}(Patten {et~al.} 2006).  The latter limit comes from available data
(J.D.~Kirkpatrick \etal in prep.) suggesting that $I$-band flux drops
dramatically for spectral types cooler than T6.  From \markcite{Vrba:04}Vrba {et~al.} (2004)
and \markcite{Golimowski:04}Golimoski {et~al.} (2004), the range T3 to T6 corresponds very roughly to
$T_{\rm eff} = 1500$ to $1100$~K.  Using a model which forms brown dwarfs
at a constant rate over 10~Gyr with power law mass functions of index 0.4
to 1.3 \markcite{Reid:02}(Reid, Gizis, \& Hawley 2002) and the theoretical models of \markcite{Burrows:03}Burrows, Sudarsky, \& Lunine (2003)
which give luminosities and $T_{\rm eff}$ as a function of brown dwarf
mass and age, we expect $3 - 5$ brown dwarfs in the IRAC shallow survey
to meet our selection criteria.  Intriguingly, there should be a similar
number of dwarfs with $T_{\rm eff} < 750$~K above the [4.5] flux limit,
although our $I < 23$ requirement would exclude them from the present
sample.

Given our desire to understand more fully the physical nature of the L/T
transition, the newly discovered T4.5 brown dwarf can serve as another
probe of the overluminosity of the early-/mid-T hump.  Its magnitudes
of $J = 16.88$ and $K_s = 16.99$ make it a difficult but not impossible
target for a dedicated near-infrared parallax program such as the on-going
one at the US Naval Observatory in Flagstaff \markcite{Vrba:04}(Vrba
{et~al.} 2004).  More importantly, \tdwarf\ is the first example
of a field T dwarf selected by mid-infrared photometry supplemented
by other ground-based optical and near-infrared data.  This implies
that a very similar selection technique to be employed by the {\it
Wide-Field Infrared Survey Explorer} \markcite{Eisenhardt:03}({\it WISE};
Eisenhardt \& Wright 2003), planned for launch in 2009, is sound and
will be capable of discovering other T dwarfs, and hopefully cooler Y
dwarfs \markcite{Kirkpatrick:03}(Kirkpatrick 2003).  {\it WISE} will
sample hundreds of times more volume than the IRAC Shallow Survey in
bands similar to [3.6] and [4.5], and should reveal whether there are
brown dwarfs closer to the Sun than Proxima Centauri.


\subsection{\qso:  $z = 6.12$ Quasar}

\begin{figure}[!t]
\begin{center}
\plotfiddle{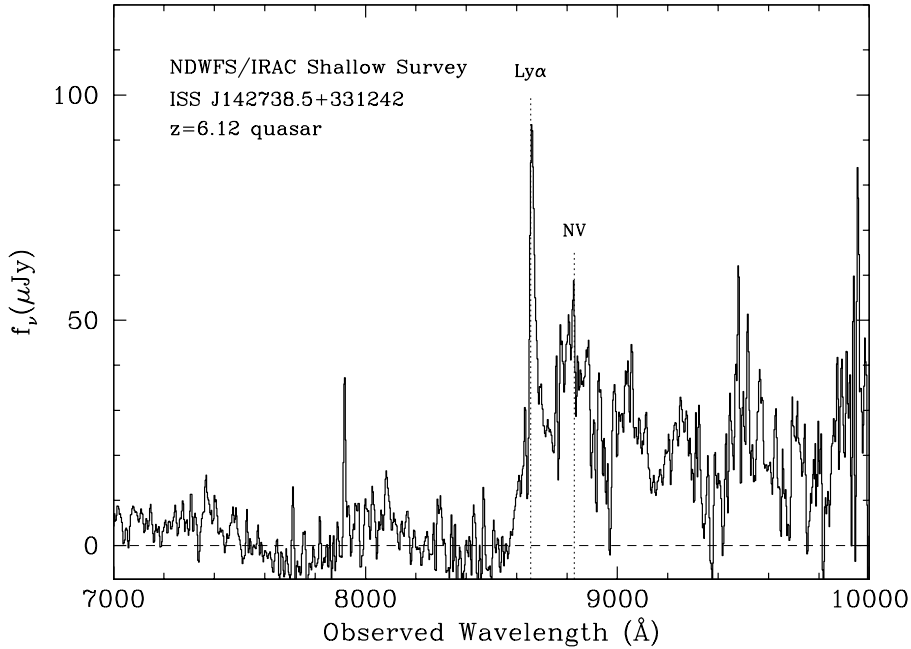}{3.1in}{-90}{45}{45}{-180}{260}
\end{center}

\caption{Spectrum of \qso, a quasar at $z = 6.12$, obtained with the
MARS spectrograph on the Kitt Peak 4m Mayall telescope.   The relative
flux calibration was determined from observations of standard stars
from the same observing run with the same instrumental configuration.
As the nights were not photometric, the spectrophotometric scale has
been estimated from the imaging.}

\label{fig.qso} 
\end{figure}

The spectrum of \qso\ (Fig.~\ref{fig.qso}) clearly shows the strong
Ly$\alpha$ emission and strong Ly$\alpha$ decrement of a $z \geq 6$
quasar.  At a redshift of $z = 6.12$, \qso\ is emitting when the universe
was 0.89~Gyr old, or only 7\%\ of its current age.  This is the tenth $z
\geq 6$ quasar identified to date, with the prior nine identified by the
Sloan Digital Sky Survey \markcite{Fan:06}(SDSS; Fan {et~al.} 2006).  \qso\ was identified
independently by \markcite{McGreer:06}McGreer {et~al.} (2006) using different selection criteria.
Two characteristics separate \qso\ from the other nine $z \geq 6$ quasars
known.  First, while the other nine were identified from 6550 deg$^2$
of the wide-area, shallow SDSS optical survey with $J$-band follow-up,
\qso\ was identified in a more sensitive, multi-wavelength survey
of only 8 deg$^2$.  Consequently, this is the least luminous quasar
known at $z \approx 6$.  Secondly, the Faint Images of the Radio Sky at
Twenty-cm survey \markcite{Becker:95}(FIRST; Becker, White, \& Helfand 1995) identifies a source with an
integrated flux of 1.03~mJy within 1 arcsec of the quasar coordinates.
\qso\ is thus the only $z \geq 6$ mJy-radio source currently known.

The evolution of the fraction of quasars which are radio loud, and,
in fact, the definition and very existence of such a dichotomy, has
been the subject of substantial literature.  Some researchers prefer a
definition based on the radio-optical ratio $R_{\rm ro}$ of the specific
fluxes at rest-frame 6~cm (5~GHz) and 4400 \AA\ \markcite{Kellerman:89}(Kellerman {et~al.} 1989).
The other common definition divides the populations at some rest-frame
radio luminosity; \eg \markcite{Gregg:96}Gregg {et~al.} (1996) uses a cutoff value for the 1.4
GHz specific luminosity, $L_{\rm 1.4~GHz} = 10^{32.5}\, h_{50}^{-2}\,
{\rm ergs}\, {\rm s}^{-1}\, {\rm Hz}^{-1}$ to separate radio loud and
radio quiet sources.\footnote{An Einstein-de~Sitter cosmology is assumed.}
The latter definition is immune to obscuration from dust, and, as argued
by \markcite{Peacock:86}Peacock, Miller, \& Longair (1986) and \markcite{Miller:90}Miller, Peacock, \& Mead (1990), is the more physically
meaningful definition.  Based on radio observations of all $z > 4$
quasars known as of mid-1999 and using the radio luminosity definition,
\markcite{Stern:00a}Stern {et~al.} (2000) found that approximately 12\%\ of quasars are radio
loud, with no evidence of this fraction depending on either redshift
(for $2 \simlt z \simlt 5$) or optical luminosity (for $-25 \simgt M_B
\simgt -28$).  For a typical radio spectral index $\alpha = -0.5$ and
an Einstein-de~Sitter cosmology for comparison with previous literature,
\qso\ has a radio luminosity of $L_{\rm 1.4~GHz} = 1.33 \times 10^{33}\,
h_{50}^{-2}\, {\rm ergs}\, {\rm s}^{-1}\, {\rm Hz}^{-1}$, classifying
it as radio loud.  \markcite{McGreer:06}McGreer {et~al.} (2006) show that this source is still
classified as radio loud based on a radio-optical ratio definition.
\qso\ is thus the most distant radio-loud quasar known.


\qso\ is only slightly fainter in luminosity than the $z \geq 6$ SDSS
quasars, so we consider all ten $z > 6$ quasars as a single sample,
deferring issues of the likelihood of our having found such a source
in our drastically smaller survey (discussed next).  The implication is
that the radio loud fraction remains near 10\%\ out to $z \approx 6.5$.
Conventional wisdom and morphological studies suggest that luminous,
radio loud AGN are preferentially identified with early-type galaxies
\markcite{McLure:99}(\eg McLure {et~al.} 1999).  Theory can explain the trend, since
early-type galaxies are likely the products of major mergers and two
coalescing supermassive black holes appear necessary to create black holes
of sufficient spin to generate highly collimated jets and powerful radio
sources \markcite{Wilson:95}(\eg Wilson \& Colbert 1995).  Assuming the radio loud -- luminous
host galaxy relation remains robust at high redshift, the apparent
discovery that $\approx 10\%$ of quasars are radio loud out to the highest
redshifts probed has interesting implications for the formation epoch of
massive galaxies. In hierarchical models of galaxy formation, late-type
(less massive) systems form first and mergers are required to form the
early-type (more massive) systems.  Eventually, therefore, one expects
the radio-loud fraction of AGNs to fall precipitously with redshift. Our
results show this epoch lies beyond $z \approx 6$, providing further
evidence for an early formation epoch for massive galaxies.  The stellar
masses of $i$-dropout galaxies in the Great Observatories Origins Deep
Survey \markcite{Giavalisco:04a}(Giavalisco {et~al.} 2004) leads to a similar conclusion from a very
different data set and line of argument \markcite{Yan:05, Yan:06, Eyles:06}(Yan {et~al.} 2005, 2006; Eyles {et~al.} 2006).


How likely was the discovery of this distant quasar in an 8 deg$^2$ field?
Interpolating the $K_s$ and $3.6~\mu$m photometry for \qso\ implies
$m_{\rm AB}[(1+z) 4400 {\rm \AA}] \approx 19.6$, or $M_{\rm AB}(4400)
= -27.2$.  For a typical quasar optical spectral index, the conversion
between AB-system $M_{\rm AB}(4400)$ and Vega-system $M_B$ is $M_B =
M_{\rm AB}(4400) + 0.12$ \markcite{Stern:00a}(\eg Stern {et~al.} 2000), implying $M_B = -27.1$
for \qso.  Our $J$-band photometry implies a continuum flux density
of $\approx 18~\mu$Jy at $1~\mu$m, or a rest-frame UV luminosity of
$M(1450) = -26.0$, making this source fainter than any of the $z \approx
6$ quasars identified by the SDSS \markcite{Fan:06}(Fan {et~al.} 2006).  \markcite{McGreer:06}McGreer {et~al.} (2006)
found a slightly brighter absolute magnitude, $M(1450) = -26.4$, likely
due to their alternate methodology whereby a quasar template fit to the
IRAC data was used to derive the rest-frame UV luminosities.

We estimate the number of high-redshift quasars expected from our
selection criteria using the \markcite{Fan:04}Fan {et~al.} (2004) high-redshift quasar
luminosity function, derived from the SDSS.  We approximate high-redshift
quasar spectra as step functions, with zero flux below redshifted
Ly$\alpha$ and a flat SED (in $f_\nu$) redward of Ly$\alpha$, and we
approximate the NDWFS $I$-band filter as a tophat function.  Our selection
criteria restrict our sensitivity to quasars at $5.5 \simlt z \simlt 6.5$.
The lower redshift limit comes from the $R - I$ color requirement,
determined from the \markcite{Richards:06}Richards {et~al.} (2006) model discussed in \S3; indeed,
the $z = 5.39$ quasar identifed by \markcite{Cool:06}Cool {et~al.} (2006) is too blue in $R -
I$ to meet our selection criteria (Fig.~\ref{fig.colcol}).  The upper
redshift limit corresponds to Ly$\alpha$ shifting out of the $I$-band
filter.  The \markcite{Fan:04}Fan {et~al.} (2004) luminosity function predicts 3.3 quasars at
$5.5 < z < 6.5$ with $I < 23$ in our 8 deg$^2$ survey.  This prediction
exactly matches the current results, though, notably, the faintest of
the high-redshift Bo\"otes quasars has $I = 22.0$, suggesting that more
quasars remain to be discovered with $22 < I < 23$ and that the faint end
slope of the high-redshift quasar luminosity function is steeper than
currently assumed.  Of the 3.3 quasars predicted at $5.5 < z < 6.5$,
only 0.3 are expected to be at $z > 6$, or, for 12\%\ of quasars being
radio-loud \markcite{Stern:00a}(Stern {et~al.} 2000), we only had a 4\%\ chance of identifying
a $z > 6$ radio-loud quasar in this survey.  While it is premature to
make strong claims from this small sample, our results imply possible
rapid evolution in the faint end of the quasar luminosity function and in
the radio loud fraction at high redshift.



\section{Summary and Future Prospects}

\begin{figure}[!t]
\begin{center}
\plotfiddle{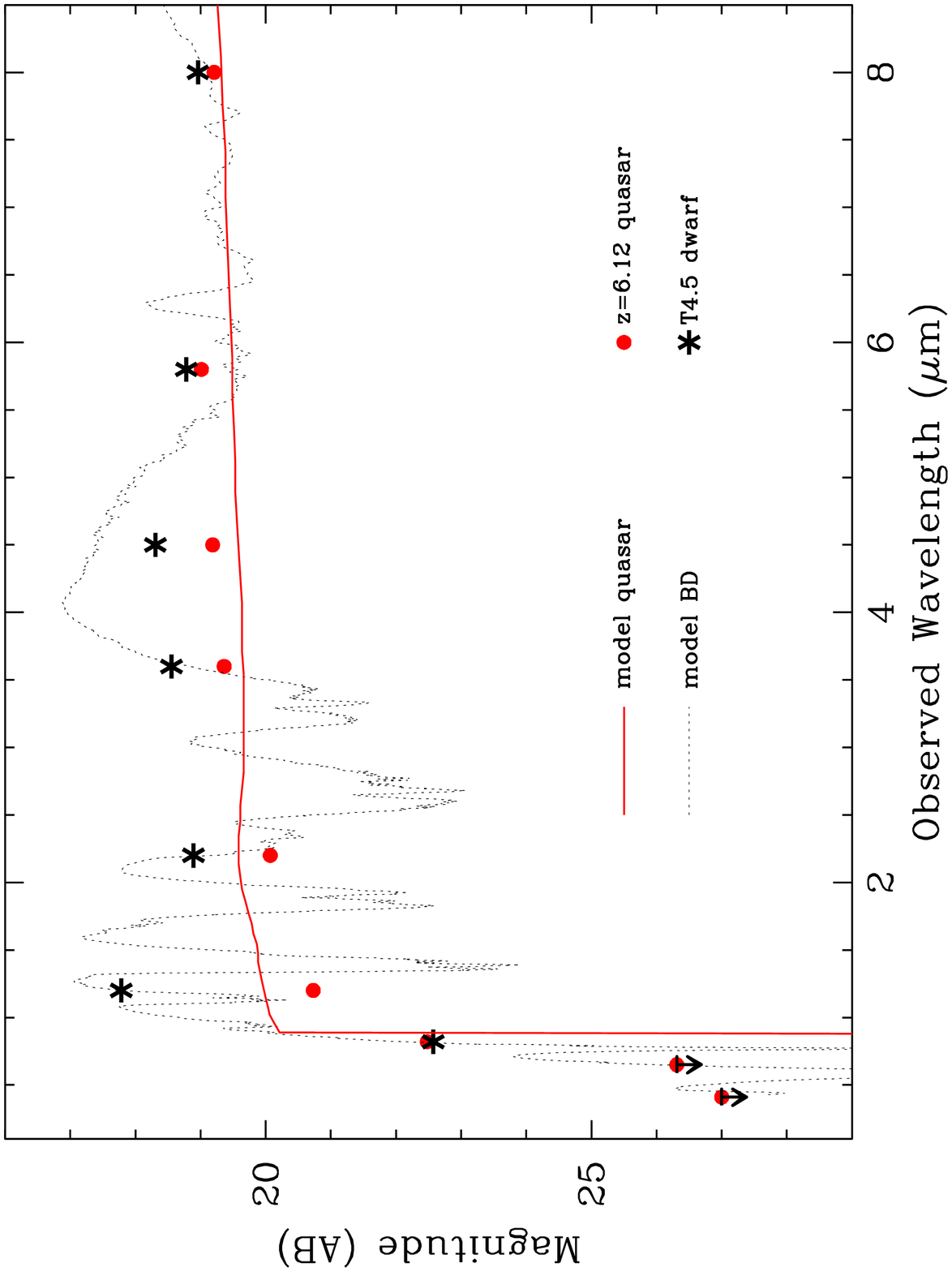}{3.1in}{-90}{45}{45}{-180}{260}
\end{center}
                                                                                                        
\caption{Spectral energy distributions of the two sources discussed in
this paper, \tdwarf, a T4.5 brown dwarf, and \qso, a $z = 6.12$ quasar.
Based on broad-band photometry, the optical and IRAC properties of these
two very different sources are nearly identical; only at near-infrared
wavelengths do they differ significantly.  The dotted line illustrates
a model brown dwarf spectrum from Burrows et al. (2006) for $T_{\rm eff}
= 1000$~K, $g = 10^5\, {\rm cm}\, {\rm s}^{-2}$, solar metallicity, and
a modal cloud particle size of $100 \mu$m.  The solid line illustrates
a model quasar at $z = 6.12$ from Richards et al. (2006), assuming no
flux is detected below redshifted Ly$\alpha$.}

\label{fig.sed}
\end{figure}

\begin{figure}[!t]
\begin{center}
\plotfiddle{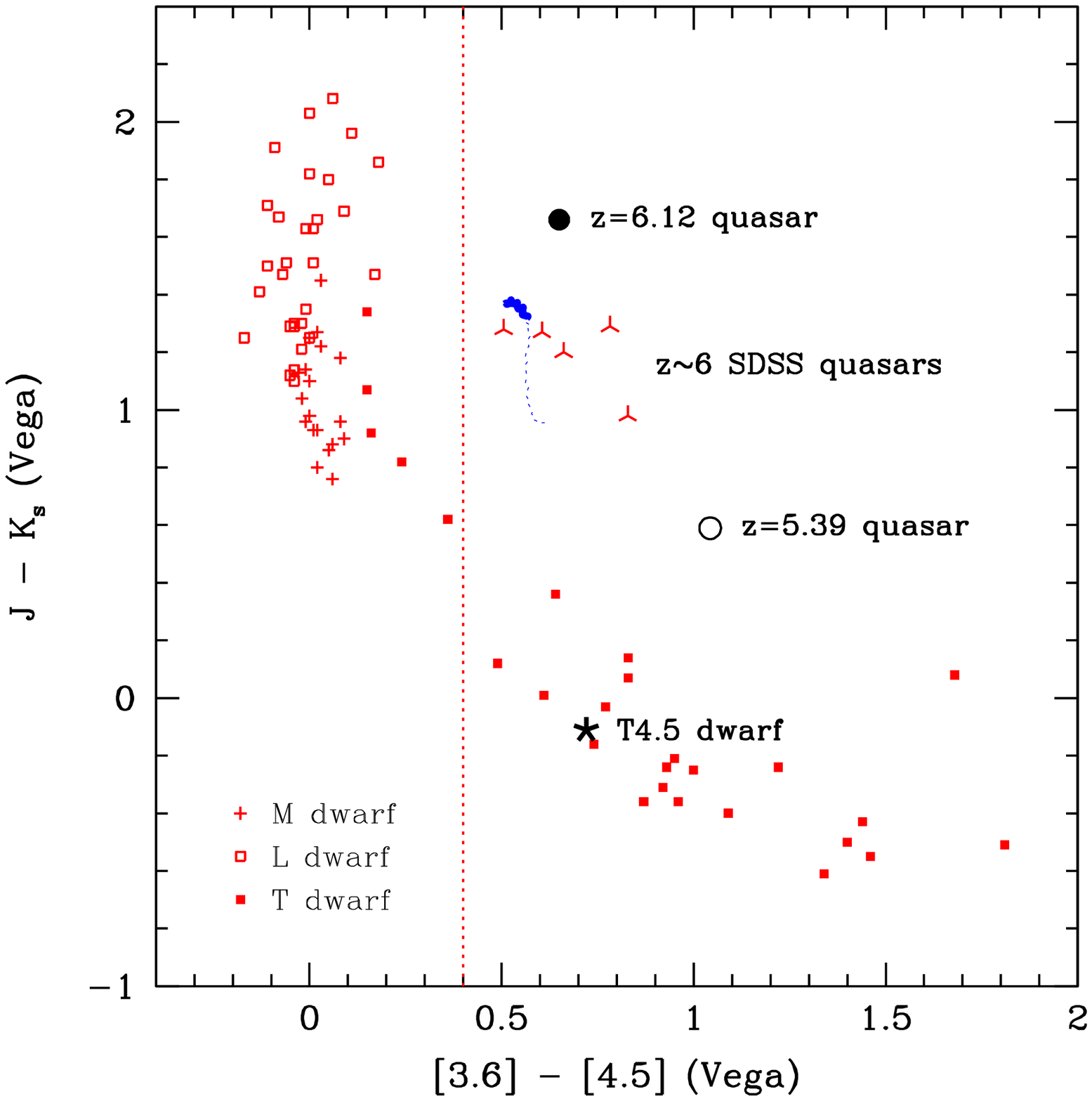}{3.0in}{0}{50}{50}{-160}{-80}
\end{center}
                                                                                        
\caption{Color-color diagrams for cool stars and high-redshift quasars.
Black symbols are for sources in the Bo\"otes field: the asterisk refers
to \tdwarf\ (T4.5 brown dwarf), the filled circle refers to \qso\ ($z =
6.12$ quasar), and the open circle refers to the $z = 5.39$ quasar in
Cool et al. (2006).  The vertical dotted line illustrates the selection
criteria employed in Cool et al. (2006) to identify luminous AGN, [3.6]
$-$ [4.5] $>$ 0.4.  Photometry of M, L, and T dwarfs from Patten et
al. (2006) are plotted in red, as indicated.  Photometry of $z \approx
6$ SDSS quasars from Jiang et al. (2006) are plotted as inverted-Y's.
The solid blue line illustrates the colors of the SDSS quasar template
from Richards et al. (2006) for $3 \leq z \leq 7$, subject to the
Madau (1995) formulation for the opacity of the intergalactic medium
as a function of redshift (the line becomes thicker for $z \geq 5.5$).
The vertical line separates high-redshift quasars and dwarfs later than
T3 from the hotter dwarfs.  The near-infrared $J - K_s$ color looks
like a promising diagnostic to separate the coolest brown dwarfs from
the high-redshift quasars.  }

\label{fig.colnearIR} 
\end{figure}

We report the discovery of the first mid-infrared selected field brown
dwarf and the discovery of the most distant radio loud source known.
Fig.~\ref{fig.sed} plots the observed SEDs of these two sources,
with a model brown dwarf spectrum from \markcite{Burrows:06}Burrows, Sudarsky, \& Hubeny (2006) and a
model high-redshift quasar from \markcite{Richards:06}Richards {et~al.} (2006).  Interestingly,
despite nine orders of magnitude difference in luminosity distance,
or nearly 20 orders of magnitude difference in luminosity, the
broad-band optical colors, the broad-band mid-infrared colors, and the
multi-wavelength brightnesses of these two extremely disparate sources
are nearly identical.  With only $B_WRI$ and IRAC photometry, there is
no possibility to separate mid-T brown dwarfs and high-redshift quasars.
As shown in Fig.~\ref{fig.colnearIR}, however, near-infrared photometry
offers the possibility to separate the two source types.  While quasars
have red $J - K_s$ colors, the latest T dwarfs have blue near-infrared
colors, $J - K_s \simlt 0.5$.

We have presented simple color criteria which very efficiently
identify astrophysically interesting sources.  Using the Bo\"otes
$4.5 \mu$m-selected catalog, the criteria presented in \S3 identify 30
potential candidates which were trimmed down to four robust candidates
after visual inspection.  The primary weakness of the criteria is that we
have only found objects at the edge of current observations, not beyond
them -- e.g., \tdwarf\ is the among the 50 coldest brown dwarfs known
and \qso\ is the 6th most distant quasar known.  To push to new territory
such as Y dwarfs and $z > 7$ quasars will require modifying the selection
criteria in \S3, and, most likely, surveying more of the celestial sphere.
In our current search, the $I \leq 23$ criterion imposed to ensure robust
morphological selection of unresolved sources is the most restrictive
requirement.  Probing deeper would allow the detection of fainter sources,
but to identify Y dwarfs and $z > 7$ quasars will likely require robust
morphological information at longer wavelengths.

We consider the effects of relaxing the selection criteria identified in
\S3.  If we retain the requirement $I \leq 23$ but drop the morphological
requirement for a point source, we obtain 434 candidates.  Going a
magnitude more deeply in $I$-band, where NDWFS photometry is still robust
but star-galaxy morphological separation fails, nearly quadruples the
number of candidates to 1598 sources.  Galaxies at $z \simgt 1$ have red
$R-I$ colors and IRAC [3.6] $-$ [4.5] $\geq 0.4$ \markcite{Stern:05b}(\eg see model
tracks in Stern {et~al.} 2005), thereby causing significant contamination.
At $z \simgt 7$, quasars will fall out of the NDWFS $I$-band, so
one might think that selecting optical dropouts with IRAC [3.6] $-$
[4.5] $\geq 0.4$ would provide efficient criteria to identify the most
distant quasars.  Unfortunately, red galaxies are again a contaminant:
there are nearly 10,000 sources in the Bo\"otes field with [3.6] $-$ [4.5]
$\geq 0.4$ and no detection in the optical passbands.  We are currently
experimenting with various schemes to trim these large samples that
arise when morphological criteria aren't available.  One possibility,
amenable to searching for cool dwarfs, is to search for sources with
blue near-infrared colors (e.g., $J - K_s < 0.5$), but red colors in
[3.6] $-$ [4.5].  The former criterion should identify both hot and cold
Galactic stars, while the latter criterion eliminates the hot stars.
Since extragalactic sources are typically redder in $J - K_s$ (e.g.,
Fig.~7 in Elston et al. 2006), these criteria should identify T dwarfs
(and colder) irrespective of morphology.  Another solution would be to
obtain better morphological measurements, as are available in the (smaller
area) Extended Groth Strip and the COSMOS surveys, or should also be
obtainable with the new generation of wide-field, near-infrared cameras.


Our observations illustrate some of the interesting sources identifiable
from wide-area mid-infrared surveys.  After {\it Spitzer} has depleted
its cryogen, expected to occur in early- to mid-2009, wide-area 3.6 and
4.5 $\mu$m surveys are likely to be an emphasis for the observatory.
Shortly thereafter, the launch of the {\it WISE} will provide full-sky,
mid-infrared images.  Such surveys, combined with the deep, complementary
optical data expected from the Panoramic Survey Telescope and Rapid
Response System \markcite{Kaiser:05}(Pan-STARRS; Kaiser {et~al.} 2005) and the Large Synoptic
Survey Telescope \markcite{Tyson:05}(LSST; Tyson {et~al.} 2005), should prove very valuable
for studying both the nearest, coldest stars and for identifying the
most distant, luminous quasars.  The former will enhance our knowledge
of star formation and Galactic structure.  The latter will probe the
first cosmic structures, the history of the intergalactic medium, and
literally expand the limits of human knowledge.

\acknowledgements 

We thank Chris Kochanek and Steve Willner for useful comments on
the manuscript.  This work is based on observations made with the
{\it Spitzer Space Telescope}, which is operated by the Jet Propulsion
Laboratory, California Institute of Technology.  Support was provided by
NASA through an award issued by JPL/Caltech.  This work also made use
of images and/or data products provided by the NDWFS, which is supported
by the National Optical Astronomy Observatory (NOAO).  NOAO is operated
by AURA, Inc., under a cooperative agreement with the National Science
Foundation.  AD and BJ are supported by NOAO.  We thank the staff
of KPNO and Keck for their expert assistance with our observations.
Research has benefited from the M, L, and T dwarf compendium housed
at {\tt http://DwarfArchives.org}.  The authors also wish to recognize
and acknowledge the very significant cultural role and reverence that
the summit of Mauna Kea has always had within the indigenous Hawaiian
community; we are most fortunate to have the opportunity to conduct
observations from this mountain.


%
\footnotesize
\begin{deluxetable}{lrrcccccccl}
\tablecaption{Photometry of Bo\"otes Field Candidates.}
\tablehead{
\colhead{Target} &
\colhead{$B_W$} &
\colhead{$R$} &
\colhead{$I$} &
\colhead{$J$} &
\colhead{$K_s$} &
\colhead{[3.6]} &
\colhead{[4.5]} &
\colhead{[5.8]} &
\colhead{[8.0]} &
\colhead{Notes}}
\startdata
\tdwarf              & $>27.1$ & $>26.1$ & 22.12 & 16.88 & 16.99 & 15.76 & 15.04 & 15.05 & 14.56 & T4.5 dwarf \nl
\qso                 & $>27.1$ & $>26.1$ & 22.03 & 19.83 & 18.17 & 16.57 & 15.92 & 15.28 & 14.81 & $z=6.12$ \nl
                     &         &         &       &       &       &       &       &       &       & \nl
ISS~J142729.6+352209 & $>27.1$ &   23.99 & 21.38 &\nodata& 18.44 & 17.33 & 16.70 &$>15.5$&$>14.8$& $z=5.53$ \nl
ISS~J142516.3+325409 & $>27.1$ &   23.76 & 21.15 &\nodata&\nodata& 17.41 & 16.77 &$>15.5$&$>14.8$& $z=5.85$ \nl
\enddata

\tablecomments{Photometry is all Vega-based, total magnitudes.
Optical photometry is from NDWFS.  Near-infrared photometry is from
FLAMEX (Elston et al. 2006).  Mid-infrared photometry is from the IRAC
Shallow Survey (Eisenhardt et al. 2004).  Non-detection limits are the
average 5$\sigma$ limits for the relevant bands across the entire field.
Catalogued FLAMEX near-infrared photometry for the $z=6.12$ quasar was 
corrupted by the bright, neighboring star.  Photometry above comes
instead from {\tt DAOPHOT} analysis of the images, using stars in the
field to model the PSF.}


\label{tab.bootes}
\end{deluxetable}
\normalsize

\clearpage
\end{document}